\documentclass[10pt,b5paper,twoside]{padeu}  %%%% do not change this line
\usepackage{epsfig,natbib_padeu}             %%%% do not change this line

\begin{document}

\title{Irradiated closed Friedmann brane-worlds}
\author{Zolt\'{a}n Keresztes\inst{1}, 
Ibolya K\'{e}p\'{\i}r\'{o}\inst{2}}
\authorrunning{Z. Keresztes, I. K\'{e}p\'{\i}r\'{o}}
\institute{\inst{1} Departments of Theoretical and Experimental Physics, University of Szeged, 6720 Szeged, D\'{o}m t\'{e}r 9, Hungary\\
	   \inst{2} Blackett Laboratory, Imperial College, Prince Consort Road, London SW7 2BW, UK
}
\email{\inst{1}zkeresztes@titan.physx.u-szeged.hu, 
       \inst{2}ibolya.kepiro@imperial.ac.uk}

%________________________________________________________________

\abstract{
We consider the evolution of a closed Friedmann brane irradiated by a bulk
black hole. Both absorption on the brane and transmission across the brane
are allowed, the latter representing a generalization over a previously
studied model. Without transmission, a critical behaviour could be observed,
when the acceleration due to radiation pressure and the deceleration
introduced by the increasing self-gravity of the brane roughly compensate
each other. We show here that increasing transmission leads to the
disappearance of the critical behaviour.
\keywords{cosmology with extra dimensions, brane-worlds, radiating black holes}
}

\maketitle

%________________________________________________________________

\section{Introduction}

According to the original idea of Kaluza and Klein, new and compact spatial
dimensions, commensurable with the Planck scale, can be introduced in the
attempt to describe various fundamental interactions. The possibility of a
non-compact extra dimension has been advanced only recently [\citet{RS}]. The
so-called brane cosmological models contain our physical world as a
hypersurface (the brane), embedded into a warped five-dimensional bulk
space-time, in which gravity acts. Its dynamics in the bulk is determined by
the Einstein-equation, however on the brane we see only a projection of this
dynamics. Consequently, on the brane, gravitational dynamics is modified as
compared with general relativity. A modified Einstein equation for the most
general case, allowing for both exotic energy in the bulk and asymmetric
embedding, was given in [\citet{Decomp}] (and for a more generic class of
models, containing induced gravity contributions in [\citet{Induced}]). The
predictions of general relativity are recovered at low energies. With
cosmological symmetries, the brane represents our observable universe.
Branes with various other symmetries were found, like an Einstein static
brane [\citet{Einbrane}], a Kantowski-Sachs type homogeneous brane [\citet
{Einbrane2}] or a G\"{o}del brane [\citet{BTs}].

Cosmological branes embedded in a bulk with radiation escaping from the
brane were studied in [\citet{LSR}], [\citet{RadBrane}], [\citet{VernonJennings}] and 
[\citet{Langlois}]. The scenario with a bulk black hole emitting Hawking
radiation (its expression being derived for closed universes with $k=1$ in 
[\citet{EHM}], [\citet{HKV}] and [\citet{GCL}]) was studied in detail in [\citet{GK}]. There, only one
black hole was considered, with its radiation completely absorbed by the
brane. (A somewhat similar model, but with $k=0$ was considered in [\citet
{Jennings}]. There a bulk black hole is placed on each side of the brane and
the radiation is completely transmitted.) 
\begin{figure}[h]
\centering\includegraphics[width=0.45\linewidth]{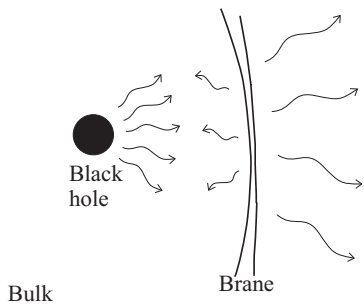}
\caption{The bulk contains an evaporating black hole. The radiation escaping
this black hole is partially absorbed, partially reflected and partially
transmitted across the brane.}
\label{Fig1}
\end{figure}

In [\citet{GK}] the case of total absorption and no transmission was analyzed
in detail. Two effects due to the radiation coming from the bulk were
identified. First, radiation pressure on the brane accelerates its outward
motion, similarly as dark energy. Second, with the increasing amount of
radiation absorbed, the self-gravity of the brane increases,
contributing towards a recollapse. However both effects represent $10^{-4}$
order perturbations for the model without radiation. (We note that the
asymmetry in the embedding produced by the presence of only one bulk black
hole behaves as an other perturbation.)

In this paper we investigate the effect of including transmission into the
model developed in [\citet{GK}], cf. Fig. \ref{Fig1} (where the possibility of
reflection is also raised). We keep a (partial) absorption on the brane, but
continue to neglect the reflection, as in [\citet{GK}], basically because there
is no known exact solution with cosmological constant describing a cross-flow
of radiation streams. Thus the five-dimensional Vaidya-anti de Sitter
(VAdS5) spacetime describes both bulk regions, the one with, and the one
without the black hole.

In order to distinguish among models with different transmissions, we
introduce a transmission rate $\varepsilon $, zero for a total absorption,
and one for a total transmission. The brane evolves cf. the energy balance,
Friedmann- and Rachaudhuri equations [\citet{Decomp}]. These equations,
specified for the case $\varepsilon \neq 0$, are given in [\citet{KKG}]. In
this paper we present some of the results of the numerical study on these
equations, for different values of the transmission rate $\varepsilon $. 

\section{Numerical results}

We assume that the brane is radiation dominated and its evolution stars at
the apparent horizon of the bulk black hole. We choose the same initial data
as in [\citet{GK}].

Allowing the transmission across the brane, we encounter new features.
First, the accumulated energy from the absorbed Hawking radiation on the
brane will be smaller than in the case of total absorption. Second, the
transmitted radiation does not contribute to the radiation pressure on the
brane, which is also smaller.

In the case with zero transmission [\citet{GK}], both the acceleration from the
radiation pressure and the deceleration from the increase in the
self-gravity of the brane were small perturbations of order $10^{-4}$, which
roughly cancel for the critical initial energy density. As in the presence
of transmission the effects are even smaller, we again will plot only 
\textit{differences}, taken in the radiating and non-radiating cases.

\begin{figure}[h]
\centering\includegraphics[width=0.6\linewidth]{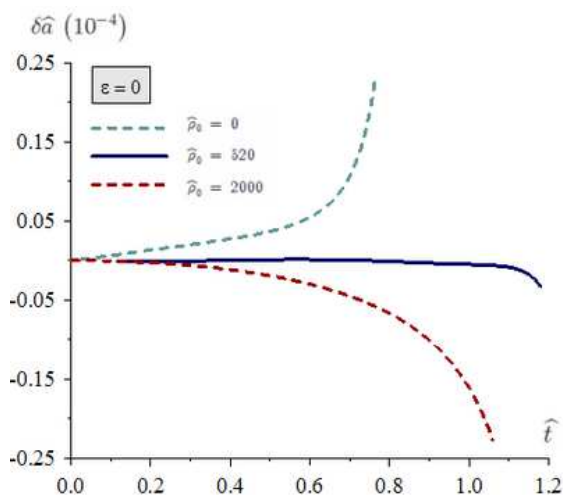}
\caption{The difference between the scale factors in the radiating and
non-radiating cases for a fully absorbing brane. The critical-like behaviour
is observed at $\widehat{\protect\rho }_{0}=520$ (solid line). For small
initial energy densities (e.g. $\widehat{\protect\rho }_{0}=0$, upper dotted
line), the pressure of the Hawking radiation is dominant. For high initial
energy densities ($\widehat{\protect\rho }_{0}=2000$, lower dotted line) the
increase of self-gravity due to absorption overtakes the radiation pressure.}
\label{Fig2}
\end{figure}

Figure \ref{Fig2} shows the evolution of the differences in the scale factor
when the radiation is switched on and off, for vanishing transmission. The
critical-like behaviour appears for $\widehat{\rho }_{0}=520$ (the quantity $
\widehat{\rho }_{0}$ denotes a properly defined [\citet{GK}] dimensionless
initial brane energy density). For lighter branes the radiation pressure is
the dominant effect, and the recollapse occurs later in the presence of
the radiation. For heavier branes the increase in self-gravity dominates and
the recollapse is speeded up.

\begin{figure}[h]
\centering\includegraphics[width=0.6\linewidth]{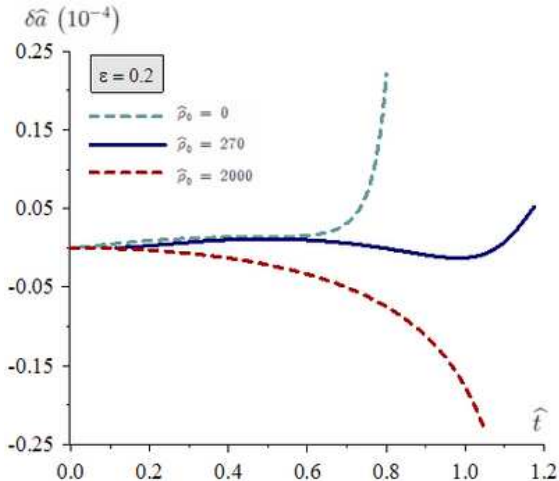}
\caption{For a higher transmission rate of $\protect\varepsilon =0.2$, the
critical brane initial energy density is less ($\widehat{\protect\rho }
_{0}^{crit}=270$) than for the case of total absorption. The amplitude
of the sinusoidal evolution of $\protect\delta \widehat{a}$ is higher.}
\label{Fig3}
\end{figure}

On Figure \ref{Fig3}, plotted for the transmission rate $0.2$ we see
similar behaviours. There are however two major differences, both due to the
appearance of the transmission. First, the critical value of the initial
brane energy density is decreased, as compared with the case of total
absorption. Second, the sinusoidal-like pattern of the critical curve is
much more accentuated, its amplitude is increased considerably.

\begin{figure}[h]
\centering\includegraphics[width=0.6\linewidth]{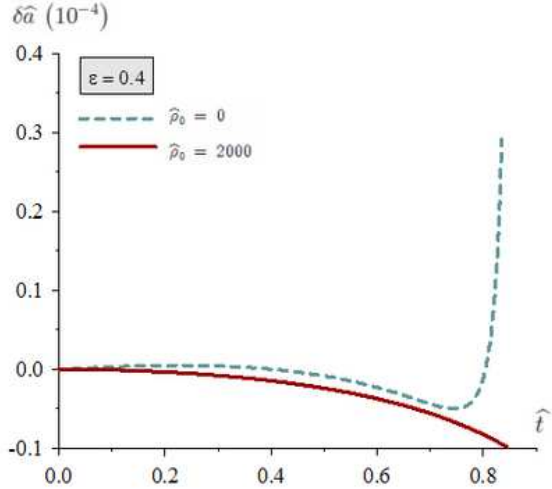}
\caption{For the transmission rate $\protect\varepsilon =0.4$ no
critical-like brane evolution can be observed. For all values of the initial
energy density the radiation drives the brane towards a faster recollapse.}
\label{Fig4}
\end{figure}

Eventually, with increasing transmission rate the critical-like behaviour
completely disappears. This is illustrated on Figure \ref{Fig4} for $
\varepsilon =0.4$.

\section{Concluding remarks}

We have considered a brane-world scenario with a brane embedded into a
bulk impregnated by radiation. The radiation is emitted by an evaporating 
bulk black hole. When this radiation hits the brane, it is partially 
absorbed and partially transmitted across the brane.

Since there is one black hole in the bulk, the embedding of the brane is
asymmetric. It was shown in [\citet{GK}], that this asymmetry is not able to
change the recollapsing fate of the closed Friedmann brane. Neither does the
completely absorbed Hawking radiation. 

In this paper, we have generalized the above statements for branes allowing 
transmission. As was the case with total absorption, with a nonvanishing 
transmission rate the absorbed radiation keeps to increase the self-gravity 
of the brane, contributing towards deceleration, while the radiation pressure 
contributes towards the acceleration of the brane. 

Our main result is that there are different behaviours depending on the 
actual value of $\varepsilon $. Whether the Hawking radiation contributes 
toward the recollapse of the brane-world universe or it speeds it up, 
depends on both the transmission rate and the brane energy density.

For small transmission rates and properly chosen initial brane energy density,
the two effects can compensate each other. We have shown that by 
increasing the transmission rate, the critical-like behaviour disappears. 
In conclusion, a high transmission rate speeds up the recollapse, 
regardless of the value of the initial brane energy density.

\section{Acknowledgement}

We thank L\'{a}szl\'{o} \'{A}rp\'{a}d Gergely for raising this problem and
guidance in its elaboration. This work was supported by OTKA grants no.
T046939 and TS044665.


\begin{thebibliography}{}

\bibitem[Barrow and Tsagas (2004)]{BTs} 
Barrow, J. D., Tsagas, C. G. 2004 \prd, 69, 064007

\bibitem[Emparan et al. (2000)]{EHM} 
Emparan, R., Horowitz, G. T., Myers, R. C. 2000 \prl, 85, 499

\bibitem[Gergely and Maartens (2002)]{Einbrane} 
Gergely, L. \'A., Maartens, R. 2002 Class. Quantum Grav., 19, 213

\bibitem[Gergely (2003)]{Decomp} 
Gergely, L. \'{A}. 2003 \prd, 68, 124011

\bibitem[Gergely (2004)]{Einbrane2} 
Gergely, L. \'A. 2004 Class. Quantum Grav., 21, 935

\bibitem[Gergely et al. (2004)]{RadBrane} 
Gergely, L. \'{A}., Leeper, E., Maartens, R. 2004 \prd, 70, 104025

\bibitem[Gergely and Maartens (2005)]{Induced} 
Gergely, L. \'{A}., Maartens, R. 2005 \prd, 71, 024032

\bibitem[Gergely and Keresztes (2006)]{GK} 
Gergely, L. \'{A}., Keresztes, Z. 2006 J. Cosmol. Astropart. Phys., JCAP01, 022

\bibitem[Guedens et al. (2002)]{GCL} 
Guedens, R., Clancy, D. and Liddle, A. R. 2002 \prd, 66, 043513

\bibitem[Hemming and Keski-Vakkuri (2001)]{HKV} 
Hemming, S., Keski-Vakkuri, E. 2001 \prd, 64, 044006

\bibitem[Jennings and Vernon (2005)]{VernonJennings} 
Jennings, D., Vernon, I. R. 2005 J. Cosmol. Astropart. Phys., JCAP07, 011

\bibitem[Jennings et al. (2005)]{Jennings} 
Jennings, D., Vernon, I. R., Davis, A.-C., van de Bruck, C. 
2005 J. Cosmol. Astropart. Phys., JCAP04, 013

\bibitem[Keresztes et al. (2006)]{KKG} 
Keresztes, Z., K\'{e}p\'{\i}r\'{o}, I., Gergely, L. \'{A}. 2006 Semi-transparent brane-worlds, in preparation

\bibitem[Langlois et al. (2002)]{LSR} 
Langlois, D., Sorbo, L., Rodr\'{\i}guez-Mart\'{\i}nez, M. 2002 \prl 89, 171301

\bibitem[Langlois (2005)]{Langlois} 
Langlois, D. 2005 [hep-th/0509231]

\bibitem[Randall and Sundrum (1999)]{RS} 
Randall, L., Sundrum, R. 1999 \prl, 83, 4690

\end{thebibliography}
\end{document}